\documentstyle[twoside,fleqn,espcrc2,epsf]{article}
\newcommand\figcaption[1]{\vskip-0.38truein\caption{#1}\vskip0.1truein}

\def\eff{{\rm eff}}
\def\MSbar{{\ifmmode\let\finish=\relax\else$\let\finish=$\fi
            \mathpalette{\hbox\bgroup$}{\overline{MS}\egroup$}%
            \finish}}
\def\mbar{\hbox{$\overline{m}$}}

\def\lilj{\hbox{$\{ U_i U_j \}$}}

\def\sslj{\hbox{$ \{ SS,\ S U_j   \}$}}

\newcommand\GeV{\mathord{\rm \;GeV}}
\newcommand\MeV{\mathord{\rm \;MeV}}

\newcommand\etal{{\it et al.}}

\begin{document}

\title{Testing the chiral behavior of the hadron spectrum\thanks{%
Poster presented 
by Tanmoy Bhattacharya and Rajan Gupta.  These calculations have been done on
the CM5 at LANL as part of the DOE HPCC Grand Challenge program, and
at NCSA under a Metacenter allocation.}}

\author{Tanmoy Bhattacharya,
        \address{T-8 Group, MS B285, Los Alamos National
        Laboratory, Los Alamos, New Mexico 87545 U.~S.~A.~} \addtocounter{address}{-1}%
        Rajan Gupta,\addressmark\ %
        and Stephen Sharpe
        \address{Physics Department, University of Washington, Seattle, WA 98195 U.~S.~A.~}
}

\begin{abstract}
We analyze the chiral behavior of the hadron spectrum obtained with
quenched Wilson fermions on 170 $32^3 \times 64$ lattices at $\beta =
6.0$.  We calculate masses of hadrons composed of both degenerate and
non-degenerate quarks. We reduce the statistical errors in
mass splittings by directly fitting to the ratio of correlation functions.
We find significant deviations from a linear dependence on the quark mass, 
deviations that are consistent with the higher order terms
predicted by quenched chiral perturbation theory.
Including these corrections yields splittings in the baryon octet
that agree with those observed experimentally.
Smaller higher order terms are also present in $m_\rho$ and $m_N$.
By contrast, the decuplet baryons are well described by a linear mass term.
We find the decuplet splittings to be $30\%$ smaller than experiment.
We extrapolate our data to $a \to 0$ by combining with the GF11 results, 
and the best fit
suggests that the quenched approximation is only good to $10-15\%$.

\end{abstract}

\maketitle

\makeatletter 

\setlength{\leftmargini}{\parindent}
\def\@listi{\leftmargin\leftmargini
            \topsep 0\p@ plus2\p@ minus2\p@\parsep 0\p@ plus\p@ minus\p@
            \itemsep \parsep}
\long\def\@maketablecaption#1#2{#1. #2\par}

\advance \parskip by 0pt plus 1pt minus 0pt

\makeatother


\section{TECHNICAL DETAILS}

We calculate quark propagators 
using the simple Wilson action and both
smeared Wuppertal and Wall sources. From these we construct three
types of hadron correlators: wall source and point sink (WL),
Wuppertal source and point sink (SL), and Wuppertal source and sink
(SS). We use five values of quark mass given by $\kappa = 0.135$
($C$), $0.153$ ($S$), $0.155$ ($U_1$), $0.1558$ ($U_2$), and $0.1563$
($U_3$), corresponding to pseudoscalar mesons of mass $2835$, $983$,
$690$, $545$ and $431$ $\MeV$ respectively where we have used
$1/a=2.33\GeV$ for the lattice scale.  We use the three light quarks
to extrapolate the data to the physical isospin symmetric light quark
mass $\mbar = (m_u+m_d)/2$. The physical value of strange quark lies
between $S$ and $U_1$ and we use these two points to interpolate to
it, and use $C$ for the charm mass.

Our overriding fitting criterion is to include as many
time-slices as possible.  We have not succeeded in developing an
automated procedure that meets this objective and works in all cases
when using two mass fits or incorporating the full covariance matrix.
For the sake of uniformity, therefore,
we use results from single mass fits keeping only
diagonal elements of the correlation matrix.  
In the cases we have checked, this does not significantly
effect either the results or error estimates.
All errors are obtained using single elimination Jackknife.

\begin{figure}[t]
\figcaption{Comparison of 
$m_{\eff}(t)$ for $U_1U_1U_1$ nucleon correlators with SL and WL sources.}
{\epsfxsize=0.9\hsize\epsfbox{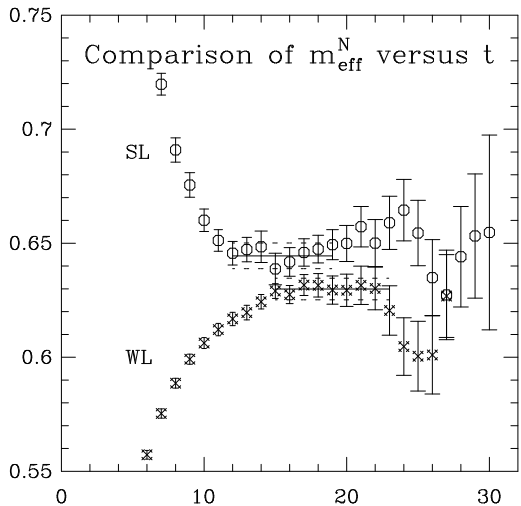}}
\label{fnuccomp}
\vskip -0.2 truein
\end{figure}

We find that masses extracted from WL correlators are
systematically lower than those from SL or SS correlators.  For the
pion and rho channels, this difference is $\sim1\sigma$, while for the
nucleons it is $\sim 2-3\sigma$ as exemplified in Fig.~\ref{fnuccomp}.
This difference arises because the signal dies out before
contamination from excited state is fully removed. In cases where the
signal persists to large enough times, the estimates from the wall and
Wuppertal sources converge to a value roughly in the middle.  Since
$SL$ and $SS$ results are highly correlated, we take as our best
estimate the average given by $(2\times WL+SL+SS)/4$. 

To extrapolate the hadron masses towards the chiral limit, and to test
the forms predicted by quenched chiral perturbation theory (CPT), we
make fits to both $1/2\kappa$ and the non-perturbative definition of
lattice quark mass described in Ref.~\cite{HM95}. The two give
virtually identical results.

\section{MESON SPECTRUM}

\begin{figure}[t]
\figcaption{A linear fit to lightest six \lilj\ points.}
{\epsfysize=0.9\hsize\epsfbox{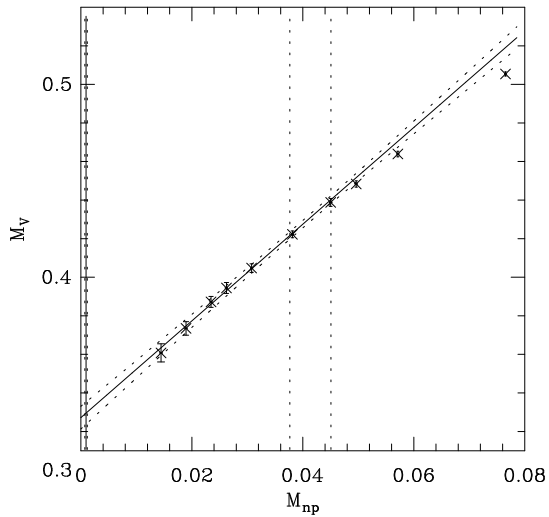}}
\label{frhovsmnp}
\vskip -0.2 truein
\end{figure}

A linear fit to $M_\pi^2$ using the six lightest \lilj\ points
gives $M_\pi^2 = 0.0013(5) + 2.296(11)m_q$. 
With our current data we cannot determine whether the non-zero 
(though tiny) intercept is due to finite size effects, 
quenched chiral logs or the absence of chiral symmetry with Wilson fermions. 
The linear fit to $M_\rho$ ($0.3296(59) + 2.54(14) m_q$) 
is shown in \ref{frhovsmnp}.  The data show clear curvature,
so we fit adding two types of corrections: $m^{3/2}$ (non-analytic term due
to chiral loops) and $m^2$. Both succeed in fitting the \lilj\ and
\sslj\ points. These fits give
\begin{eqnarray}
\kappa_c   \aftergroup\hfil &{}= 0.157131(9), \aftergroup\hfill\nonumber \\
\kappa_{\mbar} \aftergroup\hfil &{}= 0.157046(9) , \aftergroup\hfill\nonumber \\
a^{-1}(M_\rho) \aftergroup\hfil &{}= 2.330(41) {\GeV} \qquad (m \hbox{ fit}) \,, \aftergroup\hfill\nonumber \\
a^{-1}(M_\rho) \aftergroup\hfil &{}= 2.365(48) {\GeV} \qquad (m^{3/2} \hbox{ fit}) \,, \aftergroup\hfill\nonumber \\
a^{-1}(M_\rho) \aftergroup\hfil &{}= 2.344(42) {\GeV} \qquad (m^2     \hbox{ fit}) \,, \aftergroup\hfill\nonumber 
\end{eqnarray}
Since the three fits to $M_\rho$ have comparable $\chi^2$, we cannot 
study CPT in detail.

Using linear fits for pseudoscalar and vector states we determine the
lattice strange quark mass by first extrapolating $M_K^2/M_\pi^2$,
$M_{K^\ast} / M_\rho$, or $M_{\phi} / M_\rho$ to $\mbar$, and then
linearly interpolating between $U_1$ and $S$ until these quantities
match their physical values.  The results for $m_s$ ($\MSbar$ scheme,
run down to $2$ GeV) are $m_s(M_K) = 89(2) \MeV$ versus
$m_s(M_\phi)=106(6) \ MeV$, $i.e.$ a $20\%$ difference. Using
$M_K^2/M_\pi^2$ to fix $m_s$ implies that $m_s \equiv 25 \mbar$ as we
use the lowest order chiral expansion to fit the data.  On the other
hand, using $M_{\phi} / M_\rho$ gives $m_s / \mbar \approx 30$, 
in surprisingly good agreement with the next-to-leading chiral result
\cite{Donoghue}.  The differences presumably result from a combination
of quenching and discretization errors.  We use $m_{\rm s}(M_{\phi})$ here.

\section{BARYON SPECTRUM}

We calculate three types of correlation functions made up of flavors
$A, B, C$. The spin-$1/2$ $\Sigma$ type (which also includes $\Xi$ and $N$)
are labeled $A\{BC\}$ (symmetric in $B,C$);
the $\Lambda$-like are $A[BC]$ (anti-symmetric in $B,C$); and the
spin-$3/2$ decuplet are $\{ABC\}$. With 4 flavors $S,
U_i$, one can write 40 correlators of types $A\{BC\}$ and $A[BC]$
each. For $B \ne C$, SU(2) symmetry is broken and there is mixing
between the $\Lambda$ and $\Sigma$ states. 
However, as explained in \cite{HM95}, for $\delta m t << 1$ 
(as is true for our data) the mixing can be ignored.

To study mass splittings we make fits to appropriate ratios of
correlators for a given source and sink.  For example, fits to
${\Gamma_\Sigma(t) / \Gamma_N(t) } \sim e^{-(M_\Sigma - M_N) t}$, give
$(M_\Sigma - M_N$) directly. This has the advantage of both reducing
some of the sources of systematic errors and of improving the
statistical errors (the errors in $\Delta M$ are $3-5$ times smaller
than those obtained from individual fits).

Our fits are motivated by the results of quenched CPT worked out in
Ref. \cite{rSRSchiral} and are labeled chiral (non-analytic
$m_q^{3/2}$ terms) and analytic ($m_q^2$ terms). We first consider
the octet hyperfine splitting, ``$\Sigma-\Lambda$'', given by 
\begin{eqnarray}
\label{esigmlam}
& {M_{A\{BB\}} - M_{A[BB]}   \over m_A - m_B} =
(-8D/3) + c_1 { M_{AB}^3 - M_{BB}^3  \over m_A - m_B} \nonumber \\
&{} + c'_1 {M_{AA}^3 - M_{BB}^3 \over m_A-m_B} + d_1 (m_A+m_B) + d'_1
m_B \,,
\end{eqnarray}
where $M_{AB}$ is the mass of the meson with flavor $\overline{A}B$,
etc.  The constants $c_i$ and $d_i$ can be expressed in terms of
parameters of the quenched chiral Lagrangian.  For reasonable choices
of these parameters, one expects $|c'_1|\ll |c_1|$.  There is no
useful information concerning $d_1$ or $d'_i$.

\begin{figure}[t]
\figcaption{Test for chiral corrections in $M_\Sigma - M_\Lambda$.
The extrapolated value is at the extreme left.}
{\epsfxsize=0.9\hsize\epsfbox{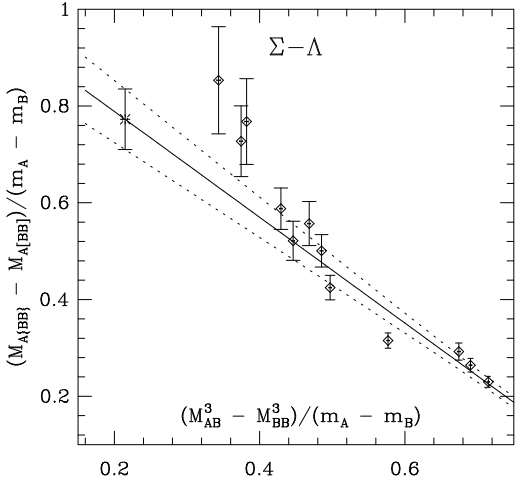}}
\label{fsiglamchiral}
\vskip -0.2 truein
\end{figure}

\begin{figure}[t]
\figcaption{Quadratic fit including baryons composed of completely non-degenerate quarks.}
{\epsfxsize=0.9\hsize\epsfbox{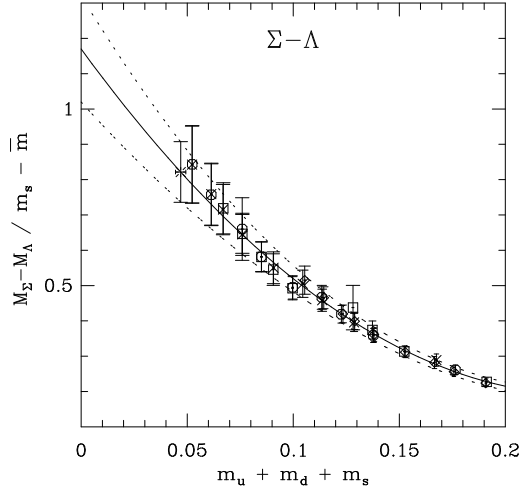}}
\label{fsiglamdiff}
\vskip -0.2 truein
\end{figure}

We fit our results in two ways. First, we assume that the $c_1$ term
is dominant, and make a linear fit (called chiral) to the 12 SU(2)
symmetric non-degenerate points as shown in Fig.~\ref{fsiglamchiral}.
The sizable non-zero slope shows that significant terms of higher 
order than linear in the quark mass are needed.
In our second fit we test to see whether our data 
can be represented as well with analytic terms. 
We find, by trial and error, that Eq.~\ref{esigmlam} with
$c_1=c'_1=0$ and $d_1=d'_1$ works, as shown in Fig.~\ref{fsiglamdiff}.
The data show significant curvature, 
so we have extrapolated to the physical point
using a quadratic fit, corresponding to $m_q^3$ terms in baryon masses!
The extrapolated value of
$M_\Sigma - M_\Lambda$ from the two fits are given in
Table~\ref{toctetdiff};  these agree with each other and with the
experimental splitting.

Next we consider the ``$\Sigma-N$'' splitting, 
\begin{eqnarray}
\label{esignuc}
&{} {M_{A\{CC\}} - M_{B\{CC\}} \over m_A - m_B} = 2(F-D)
+ c_2 { M_{AC}^3 - M_{BC}^3  \over m_A - m_B} \nonumber \\
&{} + c'_2 {M_{AA}^3 - M_{BB}^3  \over m_A - m_B} +d_2 (m_A + m_B) + d'_2 m_C \,.
\end{eqnarray}
CPT suggests that $|c'_2| < |c_2|$.
Thus we fit the data assuming $c_2$ is the dominant coefficient (chiral). 
Our best trial and error attempt with analytic corrections 
assumes $c_2=c'_2=0$ and $d_2= d'_2/4$. 
Again there is definite evidence for curvature even though 
the collapse of data on to a single curve is not as good as in the $\Sigma-\Lambda$ case.
The difference in estimates of $M_\Sigma - M_N$ given in Table~\ref{toctetdiff}
is indicative of this. 

Thirdly, we consider the difference ``$\Xi-N$'':
\begin{eqnarray}
\label{eximnuc}
&{} {M_{A\{BB\}} - M_{A\{CC\}} \over m_B - m_C} = 4F
+ c_3 { M_{AB}^3 - M_{AC}^3  \over m_B - m_C} \nonumber \\
&{} + c'_3{ M_{BB}^3 - M_{CC}^3 \over  m_B - m_C} + d_3 (m_B + m_C) + d'_3 m_A \,. 
\end{eqnarray}
Here there is {\it no} expectation that $c_3$ and $c'_3$ should be
substantially different in magnitude.  Nevertheless, our ``chiral''
fit assumes $c_3$ is the dominant coefficient, and fits the data
reasonably well.  A quadratic fit to the average quark mass (i.e.
assuming $d_3=d'_3$) is slightly better.  Again, the extrapolated
values in Table~\ref{toctetdiff} are consistent. The important point
is that in all three cases the higher order corrections are
substantial.

\begin{table}
\caption{Octet mass splittings.
Our most reliable values are from ``$m^2$'' fits.
}
\vskip 0.1 truein
\vbox{\hbox{\indent\vbox{\tabskip=0pt\offinterlineskip
\def\myskip{\omit&height1pt& && && && &\cr}
\halign {\strut#& \vrule#\tabskip=2pt&
\hfil$#$\hfil&\vrule#&
\hfil$#$&\vrule#&
\hfil$#$&\vrule#&
\hfil$#$&\vrule#\tabskip=0pt\cr\noalign{\hrule}
%
%
\myskip\myskip
&& \hbox{Fit}
&& \hbox{``$\,m^{3/2}\,$''} \hfil
&& \hbox{``$\,m^2\,$''}\hfil
&& \hbox{Expt.}\hfil
& \cr
\myskip\myskip
\noalign{\hrule}
%
%
\myskip
&& M_{\Sigma} - M_{N}     && 223(15) && 269(22) &&   253 &\cr
\myskip			                        
\noalign{\hrule}	                        
\myskip			                        
&& M_{\Xi   } - M_{N}     && 339(20) && 358(21) &&   375 &\cr
\myskip			                        
\noalign{\hrule}	                        
\myskip			                        
&& M_{\Sigma} - M_\Lambda && 78(14)  && 84(9)   &&   77 &\cr
\myskip
\noalign{\hrule}
%
%
\crcr}}}}

\label{toctetdiff}
\vskip -0.2 truein
\end{table}

Since keeping only terms linear in the quark masses provides a poor
description of most mass differences, we include an
$m_q^{3/2}$ (or $m^2$) term in our fits to the four degenerate 
combinations yielding $M_{N} = 1070(35) \MeV$ ($ 1072(31) \MeV$).
In each case we have included the same form for the
higher order corrections in $M_\rho$ when determining $a$ and $\mbar$.
For our best estimate we take the mean, $M_{N} = 1071(35) \MeV$.

\begin{figure}[t]
\figcaption{Linear fit to spin-$3/2$ baryon data. 
The four degenerate cases are shown with octagons.}
{\epsfxsize=0.9\hsize\epsfbox{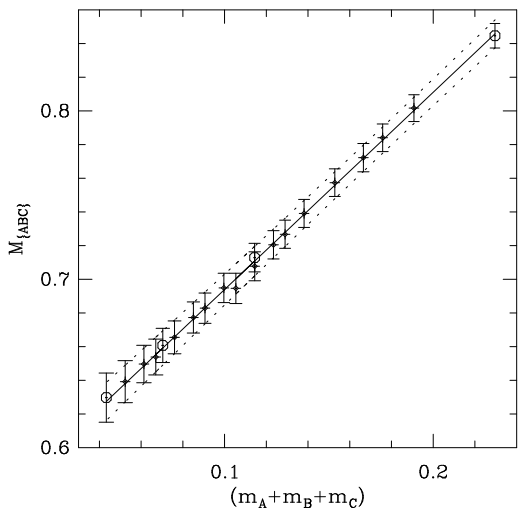}}
\label{fdecupvsmnp}
\vskip -0.2 truein
\end{figure}

In the spin-3/2 baryons we can form 20 states with our four masses,
and none of these mix with each other. As shown in
Fig.~\ref{fdecupvsmnp}, the linear fit $M_\Delta a = 0.581(13) +
1.13(5) (m_A + m_B + m_C) a $ works for all 20 quark combinations
{ABC}.  The mass differences calculated from this fit turn out to be
$\approx 25\%$ smaller than the experimental values. Note that this
discrepancy is worsened if we use $m_s(M_K)$.  Linear extrapolation of
$M_\Delta - M_N$ calculated from the ratios of correlators gives
$318(30) \MeV$.  Including an $m_q^{3/2}$ or $m_q^2$ term in both
$M_\Delta - M_N$ and $M_\rho$ fits gives $365(44)$ and $347(39)$
respectively.  These estimates are slightly higher than the
experimental value $293\ \MeV$.

\section{CONTINUUM RESULTS}

Our best estimates at $\beta=6.0$ 
with linear extrapolations in $m_q$ using only the $U_i$ quarks are 
\begin{eqnarray}
\label{emassrat}
{M_N / M_\rho} = 1.412(35) \qquad \hbox{\rm Expt: 1.22}
\,, \nonumber\\
{M_\Delta / M_\rho} = 1.800(47) \qquad \hbox{\rm Expt: 1.60}
\,, \nonumber\\
{M_\Delta / M_N}    = 1.275(36) \qquad \hbox{\rm Expt: 1.31}
\,. \nonumber
\end{eqnarray}
The GF11 collaboration \cite{GF11} has claimed, based on data at
$\beta=5.7,\ 5.93$, and $6.17$, that these ratios have a significant
slope when extrapolated to $a=0$.  We update their ``012'' and ``4''
sink fits. The change, on adding our point, to the ``012'' fit is
from $M_N / M_\rho = 1.28(7)$ to $1.30(6)$ and from $M_\Delta / M_\rho
= 1.61(8)$ to $1.62(7)$.  The $\chi^2/_{dof}$ for the new fits are
$2.1$ and $0.85$ respectively.  The analogous numbers for the sink
``4'' data are $ 1.33(9) \to 1.38(7)$ and $ 1.68(10) \to 1.73(10)$
with $\chi^2/_{dof}$ equal to $1.2$ and $0.86$ respectively.

The difference between the two fits is due to the ``012'' and
``4'' sink data at $\beta=5.7$.  If we neglect the point at strongest
coupling, $\beta=5.7$, then the remaining three points again show
smaller $a$ dependence, similar to the sink ``4'' case.  Thus, our
preferred estimates are $M_N / M_\rho = 1.38(7)$ and $M_\Delta /
M_\rho = 1.73(10)$ from sink ``4'' fit. The ambiguity in the
extrapolation makes it clear that data at more values of $\beta$ are
needed in order to reliably determine continuum results.

\end{document}